\documentclass[a4paper,fleqn,usenatbib]{mnras}
\usepackage{textcomp}
\usepackage{amsmath}
\usepackage{color}
\usepackage{amssymb}
\usepackage{graphicx}

\usepackage[T1]{fontenc}
\usepackage{aecompl}

\title[Sun et al.]{On the origin of the HLX-1 outbursts}
\author[Sun et al.]{
Mouyuan Sun,$^{1,2}$
Wei-Min Gu,$^{1,5}$\thanks{E-mail: guwm@xmu.edu.cn}
Zhen Yan,$^{3,5}$
Qingwen Wu,$^{4}$
and Tong Liu$^{1,5,6}$
\\
$^{1}$Department of Astronomy and Institute of Theoretical
Physics and Astrophysics, Xiamen University, Xiamen, Fujian 361005, China\\
$^{2}$CAS Key Laboratory for Research in Galaxies and Cosmology, Department
of Astronomy, University of Science and Technology of China, \\
Hefei, Anhui 230026, China\\
$^{3}$Key Laboratory for Research in Galaxies and Cosmology, Shanghai
Astronomical Observatory,Chinese Academy of Sciences, \\
Shanghai 200030, China\\
$^{4}$School of Physics, Huazhong University of Science and Technology, Wuhan
430074, China\\
$^{5}$SHAO-XMU Joint Center for Astrophysics, Xiamen University, Xiamen, Fujian
361005, China\\
$^{6}$Department of Physics and Astronomy, University of Nevada, Las Vegas, NV
89154, USA
}

\date{Accepted XXX. Received YYY; in original form ZZZ}

\pubyear{2016}

\begin{document}
\label{firstpage}
\pagerange{\pageref{firstpage}--\pageref{lastpage}}
\maketitle

\begin{abstract}
HLX-1, currently the best intermediate-mass black hole candidate, has undergone
seven violent outbursts, each with a peak X-ray luminosity of
$L_{\mathrm{peak},\mathrm{X}}\sim 10^{42}\ \rm{erg\ s^{-1}}$. Interestingly, the
properties of the HLX-1 outbursts evolve with time. In this work, we aim to
constrain the physical parameters of the central engine of the HLX-1 outbursts
in the framework of the black hole accretion. We find that the physical
properties of the HLX-1 outbursts are consistent with being driven by the
radiation pressure instability. This scenario can explain the evolution of the 
recurrent timescales of the HLX-1 outbursts as a function of the durations. 
\end{abstract}


\section{Introduction}
\label{sect:intro}
HLX-1, a hyper-luminous X-ray source in galaxy ESO 243-49 ($z=0.0224$), is
currently the best intermediate-mass black hole (IMBH) candidate
\citep{Farrell2009, Davis2011, Servillat2011}. Since its identification, HLX-1
has undergone seven recurrent violent X-ray outbursts, each with a peak X-ray
luminosity of $L_{\mathrm{peak},\mathrm{X}}\sim 10^{42}\ \rm{erg\ s^{-1}}$. The
recurrent timescale is roughly one year \citep{Lasota2011, Godet2014}. A detailed
analysis by \cite{Yan2015} indicates that various properties of the HLX-1
outbursts evolve with time while $L_{\mathrm{peak},\mathrm{X}}$ is roughly a
constant (see Table~\ref{tbl:t1} for more details\footnote{As $t_{\rm rec}$ is
properly measured only for the last five outbursts, we will focus on these
outbursts in the this work.}).

The hydrogen ionization instability, which is believed to drive the outbursts
in many Galactic black hole X-ray binaries (BHXRBs), is unlikely to trigger
the outbursts of HLX-1. This is simply due to the fact that the hydrogen ionization
instability is triggered only in the region of partial ionization of the hydrogen.
The corresponding viscous timescale is $\sim 100$ years \citep{Lasota2011} which
is too long for HLX-1. In addition, \cite{Yan2015} systematically quantified the
outburst properties of HLX-1, and compared their results with those of BHXRBs.
They conclude that HLX-1 does not follow the correlations that are defined by
BHXRBs \citep{YY2015}. These results indicate that, if the HLX-1 outbursts are
driven by instabilities in the accretion disk, and the typical unstable radius
is much smaller than that of the region of partial ionization of the hydrogen.

Several new mechanisms are proposed to explain the timescales. For instance,
\cite{Lasota2011} attribute the one-year recurrent timescale as the orbital
period of the donor star. That is, the outbursts are triggered as the donor
star passes the periapse of a highly eccentric orbit and overfills the Roche
lobe. However, as pointed out by \cite{Miller2014}, this model may not able to
explain the fact that both $E_{\rm rad}$ and $t_{\rm dur}$ decrease with time 
(see Table~\ref{tbl:t1}). \cite{Miller2014} 
therefore favor the wind accretion instead of the Roche lobe overflow. The evolution 
in the properties of HLX-1 outbursts is ascribed to the variability of wind velocity. 
In a word, our understanding of the HLX-1 outbursts is far from clear.

In this work, we aim to recover the physical properties of the central engine that 
drives the HLX-1 outbursts. Our methodology is introduced in
Section~\ref{sect:model}. In Section~\ref{sect:result}, we apply our model to
HLX-1. We found that the HLX-1 outbursts are consistent with being driven by
the radiation pressure instability. Discussions are made in
Section~\ref{sect:dis}.

\begin{table}
  \centering
  \scriptsize
  \caption{Properties of the HLX-1 outbursts$^a$}
  \label{tbl:t1}
  \begin{tabular}{ccccc}
    \hline
    Year of outburst & $L_{\mathrm{peak},\mathrm{X}}$ & $t_{\rm dur}$
    & $t_{\rm rec}^{b}$ & $E_{\rm rad}^{c}$\\
    years & $10^{42}\ \mathrm{erg~s^{-1}}$ & days & days & $10^{48}\ \rm{erg}$\\
    \hline
    2011 & 1.3$\pm$0.2 & 128 & 320 & $6.6\pm 0.4$ \\
    2012 & 1.4$\pm$0.4 & 110 & 334 & $5.9\pm 0.2$ \\
    2013 & 1.2$\pm$0.2 & 96 & 360 & $4.8\pm 0.2$ \\
    2014 & 1.0$\pm$0.1 & 84 & 392 & $4.3\pm 0.1$ \\
    2015 & 1.4$\pm$0.8 & 69 & 455 & $3.7\pm 0.1$ \\
    \hline
  \end{tabular}
  \\
  \scriptsize{
  $^a$ Adopted from Table 3 of \cite{Yan2015}, and for their definition, see
  Section 2.3 of \cite{Yan2015}; \\
  $^b$ The time interval between the decay phase of the current outburst and the 
  decay phase of the previous one;\\
  $^c$ The total $0.3-10\ \rm{keV}$ radiative energy during each outburst.}
\end{table}

\section{Our model}
\label{sect:model}
As pointed out by previous work \citep[e.g.,][]{Davis2011}, the central engine
of HLX-1 is most likely to be an IMBH surrounded by a cool and thin accretion 
disk. In this work, we adopted the standard Shakura-Sunyaev model \citep{ssd} 
to describe the structure of the accretion disk. In this model, the viscosity stress 
scales as the total pressure (i.e., $\propto \alpha
p_{\rm{tot}}$, where $\alpha$ is the dimensionless viscous parameter). As a
first order of approximation, we relate the properties of each outburst with
a steady time-independent disk model.  

The dimensionless accretion rate of an outburst ($\dot{m}_{\rm{bur}}$,
i.e., the ratio of the absolute accretion rate to the Eddington accretion rate,
$\dot{M}_{\rm{Edd}}=1.4\times 10^{18}M_{\rm{BH}}/M_{\odot}\ \rm{g\ s^{-1}}$)
can be roughly estimated as
\begin{equation}
  \label{eq:mdot}
  \dot{m}_{\rm{bur}} = \frac{kE_{\rm{rad}}}{t_{\rm{dur}}\eta c^2}\frac{1}{\dot{M}_{\rm{Edd}}}
\end{equation}
where $E_{\rm{rad}}$, $\eta$, $k$ and $c$ are the total radiated energy in $0.3-10\ \rm{keV}$ 
per each outburst (i.e., the integration of the X-ray luminosity over $t_{\rm{dur}}$), the 
radiative efficiency, the bolometric correction factor, and the speed of light, respectively. 

We then calculated the radial structure of the Shakura-Sunyaev disk for each $\dot{m}_{\rm{bur}}$ 
(estimated from Eq.~\ref{eq:mdot}) and obtained the corresponding radial velocity $v_{\rm{R, bur}}$ 
as a function of radius. The viscous timescale at a typical radius $R_{\rm{c}}$ is simply 
$R_{\rm{c}}/v_{\rm{R_{\rm{c}}, bur}}$. This viscous timescale controls the duration of each 
outburst, i.e., 
\begin{equation}
\label{eq:tdur}
  t_{\rm{dur}} \equiv R_{\rm{c}}/v_{\rm{R_{\rm{c}}, bur}}
\end{equation}
In principle, we can estimate $R_{\rm{c}}$ from $t_{\rm{dur}}$. 

During the quiescent state (i.e., the state between the current outburst
and the previous one), the gas supplying rate to the central IMBH (again in units of 
$\dot{M}_{\rm Edd}$) can be estimated as follows, 
\begin{equation}
\label{eq:mdotfill}
(1-f) \dot{m}_{\rm{fill}} t_{\rm{rec}} = \dot{m}_{\rm{bur}} t_{\rm{dur}}
\end{equation}
where $f$ parameterize the fraction of gas lost in e.g., outflows\footnote{For the HLX-1 outburst 
state, as the accretion rate is expected to be high, strong outflows seem inevitable \citep{GuWM2015}.} 
during each outburst.  The radial structure of the accretion disk (i.e., temperature, density, pressure, 
the radiation-to-gas pressure ratio, the radial velocity, and 
other physical properties as a function of radius) during this state was also 
calculated for each $\dot{m}_{\rm{fill}}$.

\section{Applying our model to HLX-1}
\label{sect:result}
We now estimate $R_{\rm c}$ for each HLX-1 outburst. To convert the X-ray luminosity
and energy to the bolometric ones, we assume a bolometric correction factor of 
$k=2$.\footnote{The X-ray spectral analyses of \cite{Yan2015} indicate 
that, during the luminous state, the ratio of the integrated luminosity of the disk black 
body component to $L_{\mathrm{peak},\mathrm{X}}$ is $\sim 2$ (the exact 
value depends on the inclination angle). Therefore, we adopted $k=2$. We also 
performed our calculation for $k=5$. Very similar conclusions were obtained.} We 
adopt $\eta=0.1$. The mass of the IMBH in HLX-1 is 
assumed to be $M_{\rm{BH}} = 3\times 10^4\ M_{\odot}$ \citep[i.e., the ``average'' 
value in][]{Davis2011}. For the viscosity, we consider two cases: $\alpha = 0.01$ and 
$\alpha=0.1$. For the parameter $f$ (see Section~\ref{sect:model}), we present the 
results for $f=0$ (i.e., no outflows) and $f=0.5$ (i.e., the outflow rate equals 
$\dot{m}_{\rm bur}$). For the purpose of exploring the impact of the assumed 
$M_{\rm BH}$ on our results, we also considered $M_{\rm{BH}}=10^4\ M_{\odot}$ for 
the case $\alpha=0.1$ and $f=0$. 

\begin{figure}
\centering
\includegraphics[width=\columnwidth]{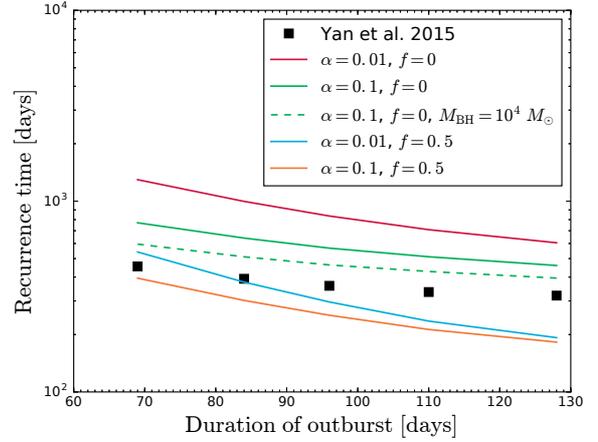}
\caption{The recurrent timescale $t_{\rm{rec}}$ as a function of the duration $t_{\rm{dur}}$. 
It is clear that $t_{\rm rec}$ is roughly consistent with $t_{\rm{vis, fill}}$ which is the viscous 
timescale at $R_{\rm c}$ in the quiescent state. Similar to that of $t_{\rm{rec}}$, 
$t_{\rm{vis, fill}}$ also decreases with $t_{\rm{dur}}$. The mass of the IMBH in HLX-1 
is assumed to be $M_{\rm{BH}} = 3\times 10^4\ M_{\odot}$, unless otherwise specified.}
\label{fig:f1}
\end{figure}

For each HLX-1 outburst, we estimated $R_{\rm c}$ via Eqs.~\ref{eq:mdot} and \ref{eq:tdur}. 
Let us take $\alpha=0.1$ and $f=0$ as an example, the expected $R_{\rm c}$ for 
the five HLX-1 outbursts are 
$69.8 \ R_{\rm S}$, $67.7 \ R_{\rm S}$, $61.9 \ R_{\rm S}$, $59.9 \ R_{\rm S}$ and 
$57.6 \ R_{\rm S}$, where $R_{\rm S}$ is the Schwarzschild radius. 
We then calculated the radiation-to-gas pressure ratio at $R_{\rm c}$ during each 
quiescent state. 
We found that this ratio is generally \textit{greater} than $1$. We therefore 
speculate that the HLX-1 outbursts are triggered within the 
radiation-pressure-dominated accretion disk. If so, it would be very interesting
to connect $t_{\rm rec}$ with the viscous timescale at $R_{\rm c}$ in the
quiescent state, $t_{\rm{vis, fill}}$. The corresponding viscous timescale 
can be estimated as $t_{\rm{vis, fill}} = R_{\rm{c}}/v_{\rm{R_{\rm{c}}, fill}}$, 
where $v_{\rm{R_{\rm{c}}, fill}}$ is the radial velocity of the accretion disk during 
the quiescent state, and the corresponding accretion rate is calculated from 
Eq.~\ref{eq:mdotfill}. 

In Figure~\ref{fig:f1}, we plot the observed time evolution of $t_{\rm rec}$.
For comparison, we also show the expected $t_{\rm{vis, fill}}$. It is clear that,
for a wide range of $\alpha$, $t_{\rm rec}$ is roughly consistent with the 
expected viscous timescale of the radiation-pressure-dominated region. 
It is also important to note that $t_{\rm{vis, fill}}$ decreases with 
increasing $t_{\rm dur}$ in way similar to that of $t_{\rm rec}$. This is due to the fact 
that $\dot{m}_{\rm{fill}}$ and therefore the radial velocity decrease with time. 

\begin{figure}
    \includegraphics[width=\columnwidth]{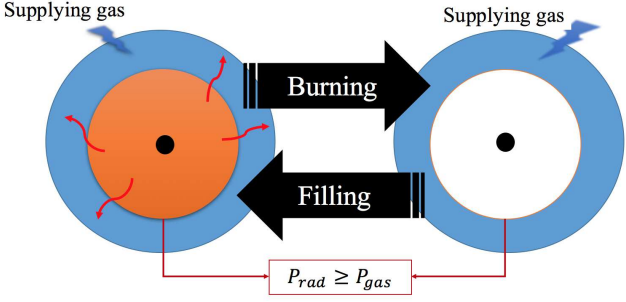}
    \caption{An illustration of our model for the HLX-1 outbursts. For sufficient
gas supplying rate, the radiation pressure dominated region (i.e., $P_{\rm{rad}}
\geq P_{\rm{gas}}$) will be unstable. As a result, HLX-1 shows quasi-periodic
outbursts. }
\label{fig:f2}
\end{figure}

Based on our results, we argue that the origin of the HLX-1 outbursts can be 
pictured as follows (see also Figure~\ref{fig:f2}). The gas supplying rate to fill 
the empty region with a typical radius of $R_{\rm c}$ is relatively high, and this 
viscous filling timescale controls the quiescent timescale between two adjacent 
outbursts. As the filling gas approaches close enough to the central IMBH, 
there would be no stable standard thin disk solution as this part is radiation 
pressure dominated instead of gas pressure 
dominated\citep[e.g.,][]{ssd,Lightman1974, Frank2002,Kato2008}. Small
temperature perturbations can result in a catastrophic growth of $\dot{m}_{\rm{bur}}$ 
(i.e., $> \dot{m}_{\rm{fill}}$) in this radiation-pressure-dominated region.
Therefore, a dramatic increase of $L_{\rm X}$ is observed. As the gas fuel in
this region is consumed, the central IMBH accretion switch offs, and leaves a
new nearly empty region around. This full cycle determines the properties of
each HLX-1 outburst. 

\section{Discussion}
\label{sect:dis}
We argued that the HLX-1 outbursts are driven by the radiation pressure
instability in the accretion disk. Such a possibility is also briefly mentioned by 
\cite{Lasota2011} although they do not perform a detailed analysis. This 
could be partially due to the fact that the radiation pressure instability itself is 
under debate. 

Indeed, although the radiation pressure instability is analytically well expected
for $L\sim 0.06\ L_{\mathrm{Edd}}$ \citep[e.g.,][]{ssd, Lightman1974, LiSL2007,
Czerny2009, XueLi2011, Zheng2011}, its observational evidence is surprisingly 
lacking as most BHXRBs remain stable for $L\sim 0.5\ L_{\mathrm{Edd}}$ 
\citep{Gierli2004}. This discrepancy has long been speculated due to the 
over-simplified description of the viscosity law \citep[e.g.,][]{Nayakshin2000,Lin2011,Zheng2011}. 
Some early magnetohydrodynamic shearing box numerical simulations 
\citep[e.g.,][]{Hirose2009}, which intend to model the viscosity self-consistently 
also find that radiation-dominated disks are thermally stable. However, recent 
new magnetohydrodynamic simulations with increased shearing box and more 
accurate radiation transfer algorithm suggest that the thermal runaway behavior 
actually exists \citep[e.g.,][]{Jiang2013}. 

Although most BHXRBs are stable against the radiation pressure instability, 
there is a notable exception, i.e., GRS 1915+105. The ``heartbeat'' like
variability in GRS 1915+105 is likely driven by the radiation pressure
instability\citep[e.g.,][]{Belloni1997}. Recently, similar ``heartbeat'' light curve is 
also discovered in IGR J17091-3624. This similarity indicates that IGR J17091-3624 
could plausibly be a second BHXRB that suffering the radiation pressure 
instability \citep{Altamirano2011}. 

In our opinion, HLX-1 could be a third one but with an IMBH. \footnote{It is true that, 
unlike the class 
$\rho$ variability of GRS 1915+105 or IGR J17091-3624, the rise timescales of the 
HLX-1 outbursts are shorter than the decline timescales. However, both GRS 
1915+105 and IGR J17091-3624 display complex variability. There are other 
variability classes (e.g., the class $\lambda$) of GRS 1915 and IGR J17091-3624 
show the temporal behaviour of fast rise and exponential decay. These variability 
classes are also likely driven by the radiation pressure instability \citep{Belloni1997}.} As the critical
accretion rate to trigger the radiation pressure instability $\propto M_{\rm BH}^{-1/8}$
\citep[e.g., Section 3.2.3 of][]{Kato2008}, such instability might be more easily observed 
for IMBHs. Note that the recurrent timescales are not expected to scale linearly 
with $M_{\rm BH}$. The physical reason are as follows. First, the radial velocity around 
$R_{\rm c}$ is inversely proportional to $M_{\rm BH}$. Second, the outer boundary of 
the radiation pressure region (in units of $R_{\rm S}$) scales positively with 
$M_{\rm BH}$ \citep[see Section 
3.2.3 of][]{Kato2008}. Indeed, $R_{\rm{c}}/R_{\rm{S}}$ we obtained for HLX-1 each 
outburst is larger than that of GRS 1915+105 \citep{Belloni1997} by roughly one order 
of magnitude. Therefore, even the IMBH in HLX-1 is only $\sim 3000$ times larger than 
the BH in GRS 1915+105, the recurrent timescales of the HLX-1 outbursts could be 
$10^5$ times longer than those of the GRS 1915+105 outbursts. 

On even larger scale, the radiation pressure instability might also
play a role in some active galactic nuclei (AGNs), e.g., some young compact
radio sources \citep{Czerny2009,WuQW2009}. It would be interesting to reveal
any possible scaling relations among these systems \citep{WuQW2016}.

\section{Summary}
In this work, we adopted the standard Shakura-Sunyaev model to constrain the physical 
properties of the central engine of HLX-1. We determined the typical size of the 
disk ($R_{\rm{c}}$) during the outbursts. We found that, even in the quiescent  
state, the radiation-to-gas pressure ratio at $R_{\rm{c}}$ is usually greater than 
$1$. We therefore proposed that the HLX-1 outbursts are driven by the radiation 
pressure instability. This scenario can explain the evolution of the 
recurrent timescales of the HLX-1 outbursts as a function of the durations. 

\section*{Acknowledgements}
We thank Wenfei Yu for helpful discussions. This work was supported by the National Basic 
Research Program of China (973 Program) under grants 2014CB845800 and 2015CB857004, 
the National Natural Science Foundation of China under grants 11573023, 11573009, 11473022, 
11403074, 11333004, 11222328, 11133005, and U1331101, the Knowledge Innovation Program of 
the Chinese Academy of Sciences, the CAS Open Research Program of Key Laboratory for the 
Structure and Evolution of Celestial Objects under grant OP201503, and the Fundamental 
Research Funds for the Central Universities under grants 20720140532 and 20720160024.

\end{document}